\documentclass[12pt, superscriptaddress]{iopart}

\usepackage{graphicx}
\usepackage{color}
\usepackage{iopams}  
\begin{document}

\title[]{The Prominent Charge-Transfer Effects of Trinuclear Complexes with Nominally High Nickel Valences}
\author{K Yamagami$^1$\footnote{Present address: Institute for Solid State Physics, University of Tokyo, Kashiwa 277-8581, Japan}, S Imada$^2$, K Yamanaka$^3$, T Yaji$^3$, A Tanaka$^4$, M Kouno$^5$, N Yoshinari$^5$, T Konno$^5$ and A Sekiyama$^1$}
\address{$^1$Division of Materials Physics, Graduate School of Engineering Science, Osaka University, Toyonaka, Osaka 560-8531, Japan}
\address{$^2$Department of Physical Sciences, Ritsumeikan University, Kusatsu, Shiga 525-0058, Japan}
\address{$^3$Synchrotron Radiation Center, Ritsumeikan University, Kusatsu, Shiga 525-0058, Japan}
\address{$^4$Department of Quantum Matter, ADSM, Hiroshima University, Higashihiroshima, Hiroshima 739-8530, Japan}
\address{$^5$Department of Chemistry, Graduate School of Science, Osaka University, Toyonaka, Osaka 560-8531, Japan}
\ead{kyamagami@issp.u-tokyo.ac.jp}

\begin{abstract}
Recently synthesized Rh-Ni trinuclear complexes hexacoordinated with sulfur ions, 3-aminopropanethiolate (apt) metalloligand [Ni$\{$Rh(apt)$_{3}\}_{2}$](NO$_{3}$)$_{n}$ ($n$ = 2, 3, 4), are found to be chemically interconvertible between the nominal Ni$^{2+}$ and Ni$^{4+}$ states.
In order to clarify the origins of their interconvertible nature and the stability of such a high oxidation state as the tetravalency from the physical point of view, we have systematically investigated the local 3$d$ electronic structures of [Ni$\{$Rh(apt)$_{3}\}_{2}$](NO$_{3}$)$_{n}$ 
by means of soft X-ray core-level absorption spectroscopy (XAS).
The experimental data have been reproduced by the single-site configuration-interaction cluster-model simulations, which indicate that the charge-transferred configurations are more stable than the nominal $d$-electron-number configuration for $n=3,4$ leading to the prominent charge-transfer effects. 
These are also supported by S $K$-edge XAS of [Ni$\{$Rh(apt)$_{3}\}_{2}$](NO$_{3}$)$_{n}$. 
Our results imply that the found charge-transfer effects have a key role to realize the interconvertible nature as well as the stability of the high oxidization state of the Ni ions.
\end{abstract}
\maketitle

\section*{Contents}
\begin{description}
  \item[1.] Introduction
  \item[2.] Methods
  \item[] Soft X-ray Absorption Spectroscopy
  \item[] Configuration-interaction cluster model
  \item[3.] Results \& Discussions
  \item[] Sulfur-coordination dependence of hybridization effects in Ni$^{2+}$ complexes
  \item[] The Prominent Charge-Transfer Effects of [Ni\{Rh(apt)$_{3}$\}$_{2}$](NO$_{3}$)$_{n}$
  \item[4.] Conclusions
  \item[] Acknowledgements
  \item[] References
\end{description}
\section{Introduction}

\begin{figure}
\begin{center}
\includegraphics[width=12cm]{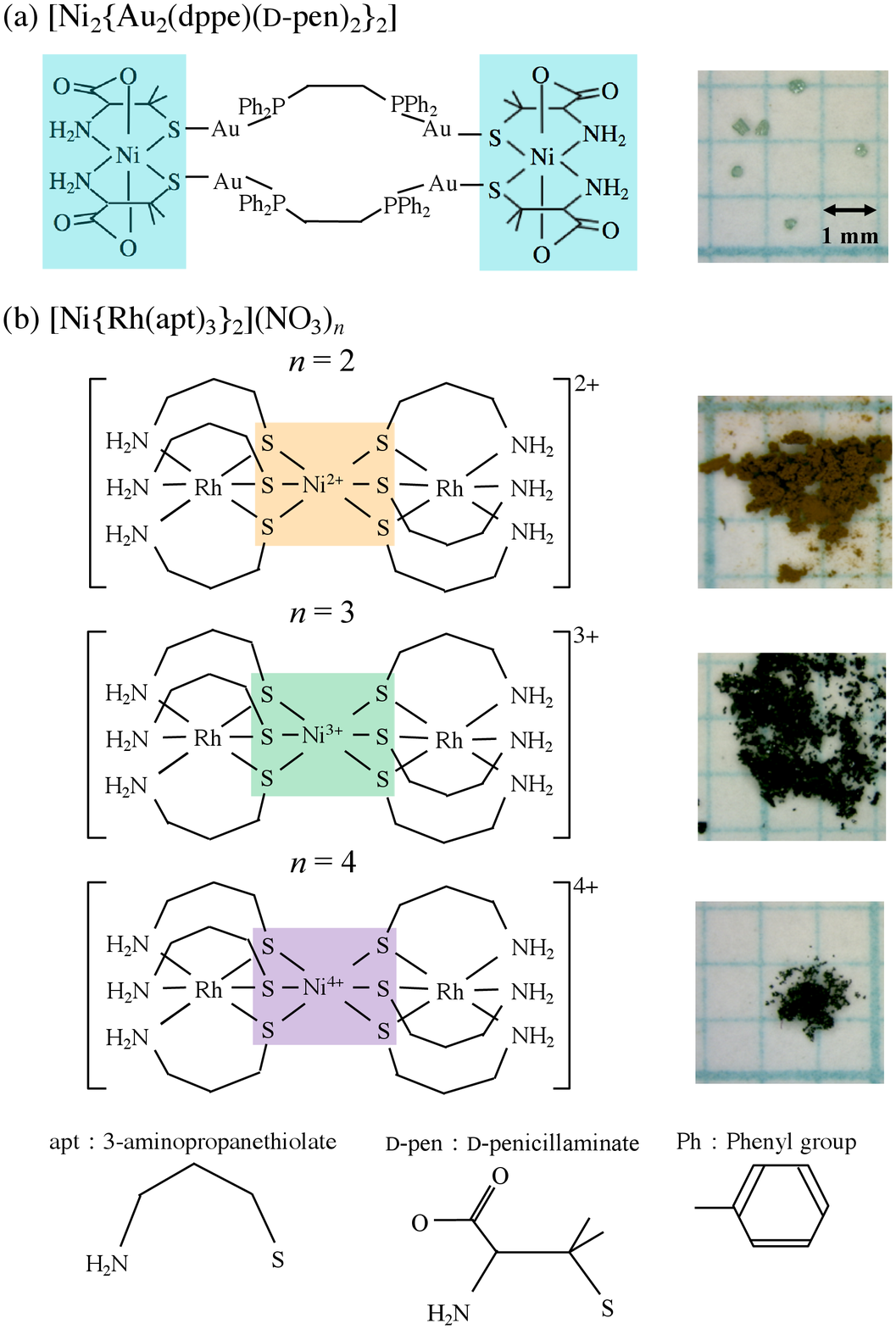}
\end{center}
\vspace{-5mm}
\begin{center}
\caption{(Color online) Molecular structures of (a) [Ni$_{2}$\{Au$_{2}$(dppe)({\scshape d}-pen)$_{2}$\}$_{2}$]~\cite{dppeII} and (b) [Ni\{Rh(apt)$_{3}$\}$_{2}$](NO$_{3}$)$_{n}$ ($n$ = 2, 3, 4)~\cite{apt}. These complexes have the sulfur two- or six-coordinating environment in octahedral-like symmetry, displayed as color marks. The size of crystal samples is less than $500\ \mu m$, as shown in the optical microscope image on right side. Molecular structures of apt, {\scshape d}-pen and phenyl group are shown in the bottom.}
\label{Fig.1}
\end{center}
\end{figure}

Sulfur-containing transition metal (TM) compounds possess a variety of fascinating properties such as metal-insulator transitions~\cite{MITI,MITII}, thermoelectric effects ~\cite{TEMI,TEMII}, and superconductivity~\cite{SCI}.
Among them, many sulfur-coordinating TM multinuclear complexes have been synthesized through stepwise synthesis, which have a potential for new physical properties~\cite{MNCI,MNCII,MNCIII,MNCIV}.
For instance, the single-crystalline [$M$$_{2}$\{Au$_{2}$(dppe)({\scshape d}-pen)$_{2}$\}$_{2}$]X$_{2}$ [dppe denotes 1,2-bis(diphenylphosphino)-ethane, $M$ = Co, Ni, and X$_{2}$ stand for anions] of which $M$ sites are unusually stabilized in an octahedral structure bonded by the coordinations with two aliphatic thiolato, two amine, and two carboxylate donors~\cite{dppeI, dppeII}, shows catalyst-like reactions for $M$ = Co~\cite{Codppe}.
On the other hand, in case of $M$ = Ni as shown in Fig.~\ref{Fig.1}(a), the chemical analysis suggests that the electronic state of Ni ions in [Ni$_{2}$\{Au$_{2}$(dppe)({\scshape d}-pen)$_{2}$\}$_{2}$] is in high-spin (HS-) Ni$^{2+}$ states~\cite{dppeII}.

Meanwhile, the nickel-rhodium trinuclear complexes with 3-aminopropanethiolate (apt) [Ni\{Rh(apt)$_{3}$\}$_{2}$](NO$_{3}$)$_{n}$ ($n$ = 2, 3, 4) have been independently synthesized [Fig.~\ref{Fig.1}(b)]~\cite{apt}.
These complexes are interconvertible through redox reaction with keeping coordination geometry around metal centers, due to an appropriate structural flexibility of [Rh(apt)$_3$] unit~\cite{apt}. 
Such an electrochemical behavior is outstanding for nickel complexes 
since the redox on the nickel centre normally induce the change of coordination geometry (square-planer for Ni$^{2+}$, octahedral for Ni$^{3+}$ and Ni$^{4+}$)~\cite{Atkins}. 
In addition, such inter-convertibility from the nominal Ni$^{2+}$ to Ni$^{4+}$ states is nor seen in inorganic Ni compounds.
The Ni site in these complexes has an octahedral-like structure with six sulfur ions.
The chemical analysis suggests that the Ni ions in [Ni\{Rh(apt)$_{3}$\}$_{2}$](NO$_{3}$)$_{n}$ are stable as a HS-Ni$^{2+}$ state for $n$ = 2, a low-spin (LS-)Ni$^{3+}$ state for $n$ = 3, and a LS-Ni$^{4+}$ state for $n$ = 4, respectively.
However, the crystal color of $n$ = 2 is different from that of [Ni$_{2}$\{Au$_{2}$(dppe)({\scshape d}-pen)$_{2}$\}$_{2}$], as shown in Fig.~\ref{Fig.1}, which implies the different electronic states of Ni$^{2+}$ ions between both complexes due to the coordination number of sulfur ions.

Conventionally, it is considered that the hybridization effects between the TM and the S atoms have a key role for determining the electronic states because of the fact that the radial distribution of S 3$p$ orbital is longer than that of O and N 2$p$ orbitals.
Actually, an importance of the Ni 3$d$-S 3$p$ hybridization effects have been reported for inorganic nickel sulfides from Ni 2$p$ core-level photoemission and absorption~\cite{NiSPES, NiSXAS, NiS2PES}.
On the other hand, in sulfur square coordinating Ni$^{n+}$ dithiolene systems, the hybridization effects have been investigated by Ni $L_{2,3}$-edge and S $K$-edge X-ray absorption spectroscopy (XAS)~\cite{Ni L-edge, S K-edgeI, S K-edgeII}.
However, the origins of the chemically interconvertible nature and stability of the high oxidation Ni$^{4+}$ state of [Ni\{Rh(apt)$_{3}$\}$_{2}$](NO$_{3}$)$_{n}$ are unclear.
The fundamental informations of the local electronic states around the Ni ions from a physical point of view would help to reveal the chemical nature of these complexes.

In this paper, we report the local 3$d$ electronic states of the Ni-Rh trinuclear complexes [Ni\{Rh(apt)$_{3}$\}$_{2}$](NO$_{3}$)$_{n}$ and [Ni$_{2}$\{Au$_{2}$(dppe)({\scshape d}-pen)$_{2}$\}$_{2}$] probed by Ni $L_{2,3}$-edge and S $K$-edge XAS.
It is found that the charge-transfer effects of Ni ions are larger for [Ni$\{$Rh(apt)$_{3}\}_{2}$](NO$_{3}$)$_{n}$ than that for [Ni$_{2}$\{Au$_{2}$(dppe)({\scshape d}-pen)$_{2}$\}$_{2}$] due to the difference of sulfur coordination number.
The single-site configuration-interaction calculations on a [NiS$_{6}$]$^{n-12}$ octahedral cluster model taking the full multiplet theory into account indicate that the degree of charge-transfer effects of [Ni$\{$Rh(apt)$_{3}\}_{2}$](NO$_{3}$)$_{n}$ are larger than that of inorganic compounds.
In addition, the charge-transfer effects become stronger with increasing $n$, 
which has also been supported by the complementary result of different spectral weight of the singly occupied molecular orbital (SOMO) in the S $K$-edge XAS spectra originating form the hybridizations between the Ni and S electronic states. 
The charge-transfer effects which we have found, in other words the charge delocalization centred at the Ni sites, are limited up to the nearest S ions, which may lead to the interconvertible nature of [Ni$\{$Rh(apt)$_{3}\}_{2}$](NO$_{3}$)$_{n}$ in which the coordination structure as well as the high oxidization states are stabilized. 

\section{Methods}
\subsection{Soft X-ray absorption spectroscopy}

The Ni $L_{2,3}$-edge XAS measurements were carried out at BL-11 of Synchrotron Radiation (SR) Center in Ritsumeikan University, Japan~\cite{My paper}. 
In this beamline, so-called varied-line-spacing plane gratings were employed, supplying monochromatic photons with $h\nu$ = 40 $-$ 1000 eV.
The Ni $L_{2,3}$-edge XAS spectra ($h\nu$ = 820 $-$ 880 eV) were taken in the partial electron yield (PEY) modes with an energy resolution of $\sim$600 meV.
In the PEY mode, the micro channel plate (MCP) detecting the Auger and secondary electrons was set in the 45$^{\circ}$-depression to the photon propagation.
We applied the voltage of $-$550 V to the Au mesh, installed in the front of MCP, for the Ni $L_{2,3}$-edge measurements in order to suppress a strong background caused by the C, N and O $K$-edge absorptions in the spectra.
The S $K$-edge XAS measurements were performed at BL-10 of SR Center in Ritsumeikan University, Japan.
The available photon energy covers from about 1000 to 4000 eV by exchanging a pair of monochromating crystals~\cite{SRBL10}.
The S $K$-edge XAS spectra ($h\nu$ = 2450 $-$ 2510 eV) were taken in the total electron yield (TEY) modes with a photon energy resolution of $\sim$1 eV.
Photon energy was calibrated by the top of the Ni $L_{3}$-edge peak of NiO ($854.0$ eV), and the S $K$-edge peak of K$_{2}$SO$_{4}$ ($2481.7$ eV).
All measurements were performed at room temperature.

Highly insulating [Ni\{Rh(apt)$_{3}$\}$_{2}$](NO$_{3}$)$_{n}$ and [Ni$_{2}$\{Au$_{2}$(dppe)({\scshape d}-pen)$_{2}$\}$_{2}$] were synthesized as previously described elsewhere~\cite{dppeII, apt}.
These micro-crystal samples were thinly expanded on the conductive carbon tape attached on the sample holder in air before transferring them into the vacuum chamber.
We repeatedly measured the spectra on the same and different sample positions, confirming the data reproducibility with neither serious radiation damage nor sample-position dependence of the Ni $L_{2,3}$-edge and the S $K$-edge XAS spectra.

\subsection{Configuration-interaction cluster model}

As mentioned above, [Ni\{Rh(apt)$_{3}$\}$_{2}$](NO$_{3}$)$_{n}$ ([Ni$_{2}$\{Au$_{2}$(dppe)({\scshape d}-pen)$_{2}$\}$_{2}$]) has sulfur six- (two-)coordination under the octahedral-like symmetry, as shown in Fig.~\ref{Fig.1}.
The distances between Ni ions are longer than 7 \AA\ in all complexes~\cite{dppeII, apt}, therefore, direct electronic and magnetic interactions between the Ni sites can be negligible.
Furthermore, all samples are insulating and a para- or non-magnetic at room temperature observed from the magnetic susceptibility~\cite{apt}.
Therefore, the configuration-interaction (CI) cluster model~\cite{NiSPES}, which has been employed for many inorganic insulating transition-metal compounds, should be a good starting point for describing the local Ni 3$d$ electronic states.

We have performed Ni $L_{2,3}$-edge XAS spectral simulations of [NiS$_{6}$]$^{n-12}$ CI cluster model under the $O_{h}$ symmetry using the XTLS 9.0 program~\cite{Xtls}.
All intra-atomic parameters such as the 3$d$-3$d$ and 2$p$-3$d$ Coulomb and exchange interactions (Slater integrals, $F^{k}$ and $G^{k}$), and the 2$p$ and 3$d$ spin-orbit couplings (SOCs) have been obtained using Cowan's code~\cite{Cowan} based on the Hartree-Fock method.
The Slater integrals and SOCs were reduced to 80 $\%$ and 99 $\%$ for all spectral simulations to reproduce the Ni $L_{2,3}$-edge XAS spectra~\cite{R factorI,R factorII}.
In the cluster model, we consider the crystal-field splitting: 10$Dq$ and charge-transfer effects explicitly.
The 10$Dq$ and on-site Ni 3$d$ Coulomb interaction: $U_{dd}$ were set to $1.0$ eV and $5.0$ eV~\cite{NiSPES,Nidihalides}, respectively.
In the Ni 2$p$-3$d$ absorption final states, the Ni 2$p$ core hole-Ni 3$d$ electrons Coulomb interaction: $U_{cd}$ is also taken into account.
Conventionally, the ratio $U_{dd}/U_{cd}$ should be 0.7$-$0.9~\cite{NiS2PES, Udd_UcdI, Udd_UcdII}.
In the present study, $U_{dd}/U_{cd}$ was assumed to be 0.7.
The transfer integrals: $T_{\sigma}, T_{\pi}$ corresponding to hybridization strength between the ionic nominal configuration ($d^{m}$) and a charge-transferred configuration ($d^{m+1}\underbar{L}$) ($d$ and $\underbar{L}$ denote a $d$ electron of the TM ions and a $p$ hole of ligands respectively) are expressed in terms of Slater-Koster parameters: ($pd\sigma$) and ($pd\pi$), namely, $T_{\sigma}=\sqrt{3}(pd\sigma), T_{\pi}=2(pd\pi)$ under the $O_{h}$ symmetry~\cite{Slater-Koster}.
The ratio ($pd\sigma$)/($pd\pi$) was fixed to be $-$2.2~\cite{RatioI, RatioII, RatioIII}.
The charge transfer energy: $\Delta$ reflecting the energy benefits of an electron from the S 3$p$ to Ni 3$d$ orbitals was defined as a positive.

To reproduce the experimental XAS spectra, the configuration dependence: $\sqrt{\alpha}$~\cite{Configuration} and the final-states effect: $\beta$~\cite{Final} of the off-diagonal matrix elements are multiplied as $\sqrt{\alpha}$ = $\sqrt{1.5}$ and $\beta$ = $0.75$, respectively.
The former is originating from the different overlapping between the Ni 3$d$ and S 3$p$ orbitals due to the electronic configuration, while the latter is due to the presence of Ni 2$p$ core hole in the final states.
The ligand bandwidth was approximated as zero in this study.
Consequently, we have reproduced the Ni $L_{2,3}$-edge XAS spectra by optimizing $\Delta$ and ($pd\sigma$).
The actual local symmetry of the Ni$^{n+}$ site is lower than the $O_{h}$ symmetry for all complexes.
However, the physical parameters in the $O_{h}$ symmetry are crucial for determining the local 3$d$ electronic states of the Ni$^{n+}$ ions.
Therefore, we consider that the simulations in the $O_{h}$ symmetry would give the significant informations of the Ni 3$d$ electronic states.

\section{Results and Discussions}
\begin{figure}
\begin{center}
\includegraphics[width=15cm]{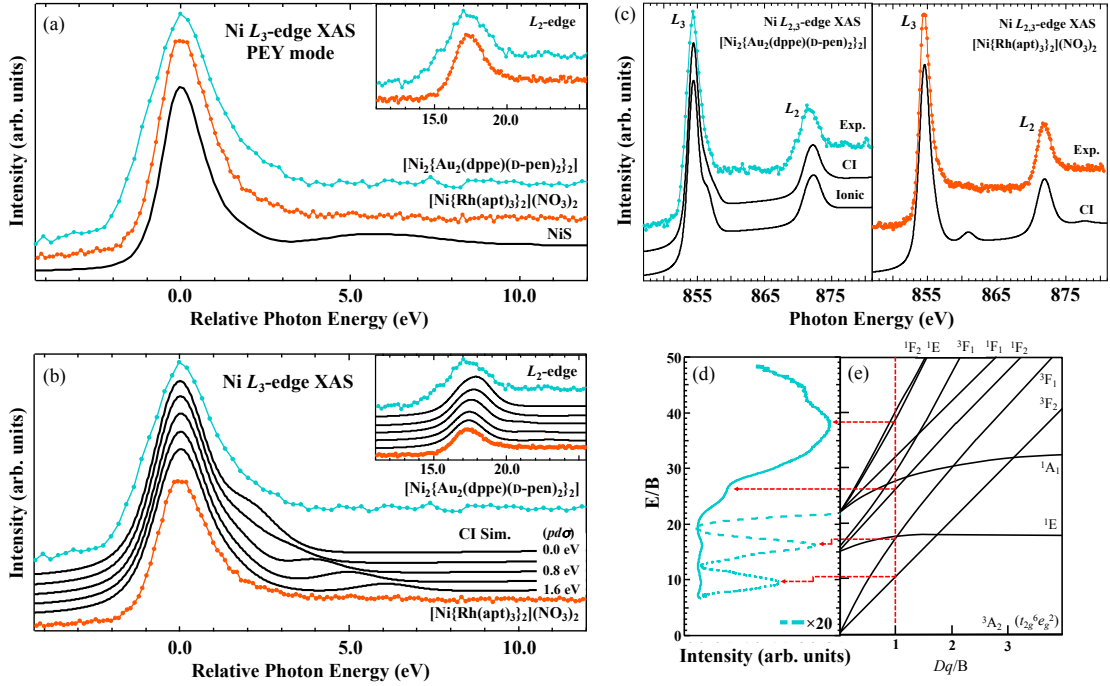}
\end{center}
\vspace{-5mm}
\begin{center}
\caption{(Color online) (a) Comparison of Ni $L_{3}$-edge XAS spectra of [Ni$_{2}$\{Au$_{2}$(dppe)({\scshape d}-pen)$_{2}$\}$_{2}$], [Ni\{Rh(apt)$_{3}$\}$_{2}$](NO$_{3}$)$_{2}$ (color solid lines with circles), and NiS (HS-Ni$^{2+}$: black solid line) cited from Ref.\cite{NiSXAS}, as a function of the relative photon energy from the peak top of $L_{3}$-edge. In all spectra, the linear background determined from the raw spectral weight in the region of $h\nu$ = $825-840$ eV are subtracted. The inset shows the Ni $L_{2}$-edge XAS spectra. (b) ($pd\sigma$) dependence of the CI cluster-model simulated (black solid lines) Ni $L_{3}$-edge XAS spectra at room temperature as a function of the relative photon energy compared with the spectra of the Ni$^{2+}$ complexes. $\Delta$ was set to 3.5 eV in these simulations. The Ni $L_{2}$-edge XAS spectra are shown in the inset. (c) Comparison between the experimental and CI cluster-model simulated (black solid lines) Ni $L_{2,3}$-edge XAS spectra for [Ni$_{2}$\{Au$_{2}$(dppe)({\scshape d}-pen)$_{2}$\}$_{2}$] and [Ni\{Rh(apt)$_{3}$\}$_{2}$](NO$_{3}$)$_{2}$ at room temperature. Optimized parameters were summarized in Table~\ref{Parameter}. In addition, simulated (black solid lines) Ni $L_{2,3}$-edge XAS spectra based on the ion model with 10$Dq$ = 1.3 eV at room temperature is shown for [Ni$_{2}$\{Au$_{2}$(dppe)({\scshape d}-pen)$_{2}$\}$_{2}$]. (d) The optical absorption spectra of [Ni$_{2}$\{Au$_{2}$(dppe)({\scshape d}-pen)$_{2}$\}$_{2}$] at room temperature cited from Ref.\cite{dppeII}. The vertical line shows the value of ${\rm E/B}$, where ${\rm E}$ and ${\rm B}$ indicate the photon energy and the Racah parameter, connected with (e) Tanabe-Sugano diagram for a $d^{8}$ configuration~\cite{TSdiagram}. The vertical red dashed line represents the value of the $Dq/{\rm B}$ obtained from 10$Dq$ = 1.3 eV. The term symbols characterized by the one electron transition excitation can correspond to the feature in (d), as connected with red dashed arrows.}
\label{Fig.2}
\end{center}
\end{figure}

\subsection{Sulfur-coordination dependence of hybridization effects in Ni$^{2+}$ complexes}
Here, we discuss the sulfur-coordination dependence of hybridization effects in Ni$^{2+}$ complexes [Ni$_{2}$\{Au$_{2}$(dppe)({\scshape d}-pen)$_{2}$\}$_{2}$] and [Ni\{Rh(apt)$_{3}$\}$_{2}$](NO$_{3}$)$_{2}$.
Figure~\ref{Fig.2}(a) shows the Ni $L_{2,3}$-edge XAS spectra of [Ni$_{2}$\{Au$_{2}$(dppe)({\scshape d}-pen)$_{2}$\}$_{2}$] and [Ni\{Rh(apt)$_{3}$\}$_{2}$](NO$_{3}$)$_{2}$ as a function of the relative energy based on the peak top of the $L_{3}$-edge.
The double-peak structure of the $L_{3}$- and $L_{2}$-edges is seen due to the Ni 2$p$ core-hole SOC in both spectra.
The main peaks of the $L_{2,3}$ edges of [Ni\{Rh(apt)$_{3}$\}$_{2}$](NO$_{3}$)$_{2}$ become narrower than those of [Ni$_{2}$\{Au$_{2}$(dppe)({\scshape d}-pen)$_{2}$\}$_{2}$].
The comparison with the XAS spectra of NiS, from which an importance of hybridization effects has been reported~\cite{NiSXAS}, suggests that the narrowing of the peaks for [Ni\{Rh(apt)$_{3}$\}$_{2}$](NO$_{3}$)$_{2}$ is ascribed to the hybridization effects.
The full width at half maximum is $\sim$2.5 eV for [Ni$_{2}$\{Au$_{2}$(dppe)({\scshape d}-pen)$_{2}$\}$_{2}$], $\sim$1.7 eV for [Ni\{Rh(apt)$_{3}$\}$_{2}$](NO$_{3}$)$_{2}$, and $\sim$1.6 eV for NiS respectively, being larger value than an energy resolution of $\sim$0.6 eV.
Actually, the Ni $L_{2,3}$-edge XAS study of Ni$^{2+}$ dihalides has demonstrated that the strong hybridization effects contribute to the narrow main peak structure due to the degradation of the intra-atomic multiplet structure, which has been explained by an Anderson model~\cite{Nidihalides}.
However, a possible satellite structure observed for NiS in the relative photon energy of $\sim$6 eV higher than the main peaks could be subtly seen at $\sim$860 eV for [Ni\{Rh(apt)$_{3}$\}$_{2}$](NO$_{3}$)$_{2}$ as shown in Fig.~\ref{Fig.2}(c) whereas a metal-like asymmetric tail~\cite{asymmetryI, asymmetryII} of the Ni $L_{2,3}$-edge is not also seen.
A satellite structure indicates the evidence of hybridization effects between the Ni 3$d$ and S 3$p$ orbitals as reported for inorganic TM compounds~\cite{NiSXAS, satelliteI}.

To confirm the hybridization effects in the spectra, the simulated Ni $L_{2,3}$-edge XAS spectra by the CI cluster-model are shown in Fig.~\ref{Fig.2}(b) as a function of ($pd\sigma$), where the $\Delta$ was fixed to 3.5 eV in these simulations.
The energy difference between the main and satellite peaks increases with ($pd\sigma$).
Note that the energy difference is mostly independent of $\Delta$ (not shown here, but a similar result is also shown in Ref.~\cite{NiSXAS}).
On the other hand, the $L_{2,3}$-edge main peaks become narrower with increasing ($pd\sigma$).
Therefore, we conclude that ($pd\sigma$) of [Ni\{Rh(apt)$_{3}$\}$_{2}$](NO$_{3}$)$_{2}$ is larger than that of [Ni$_{2}$\{Au$_{2}$(dppe)({\scshape d}-pen)$_{2}$\}$_{2}$], namely, the hybridization effects are more important for [Ni\{Rh(apt)$_{3}$\}$_{2}$](NO$_{3}$)$_{2}$, which can be caused by the different number of the Ni-S bonding and the Ni-S bonding length such as 2.454 \AA\ for [Ni$_{2}$\{Au$_{2}$(dppe)({\scshape d}-pen)$_{2}$\}$_{2}$] and 2.418 \AA\ for [Ni\{Rh(apt)$_{3}$\}$_{2}$](NO$_{3}$)$_{2}$~\cite{dppeII,apt}.

\begin{table}
 \caption{Values of $\Delta$, ($pd\sigma$) in units of eV for the CI cluster-model calculations. The effective 3$d$ electron number: $N_{d}$ and the spin state of Ni ions in initial state are determined by the simulated results. Gaussian broadening width was set to about 600 meV, meanwhile Lorentzian broadening width of $L_{3}$- ($L_{2}$-)edge was set to $\approx$ 450 meV ($\approx$ 800 meV) to reproduce the experimental Ni $L_{2,3}$-edge XAS spectra. The fraction of $d^{10-n}$, $d^{11-n}\underbar{L}$, and $d^{12-n}\underbar{L}^{2}$ in units of \% obtained from the CI cluster-model calculations for [Ni\{Rh(apt)$_{3}$\}$_{2}$](NO$_{3}$)$_{n}$ are also shown.}
  \vspace{5mm}
  \begin{tabular}{cccccccccc}
    \hline
    \hline
    &$\Delta$&($pd\sigma$)&$N_{d}$&spin&$d^{10-n}$&$d^{11-n}\underbar{L}$&$d^{12-n}\underbar{L}^{2}$\\
    \hline
    [Ni$_{2}$\{Au$_{2}$(dppe)({\scshape d}-pen)$_{2}$\}$_{2}$]&3.5&0.4&8.03&HS&96.8&3.2&0.0&\\
     \ [Ni\{Rh(apt)$_{3}$\}$_{2}$](NO$_{3}$)$_{n}$\\
    $n$ = 2&3.5&1.7&8.33&HS&69.0&28.5&2.5\\
    $n$ = 3&1.0&1.8&7.93&LS&28.4&52.2&17.8\\
    $n$ = 4&$-$4.0&0.9&7.59&LS&4.5&41.6&44.6\\
    \hline
    \hline
  \label{Parameter}
  \end{tabular}
\end{table}
The experimental Ni $L_{2,3}$-edge XAS spectra compared with the simulations of [Ni$_{2}$\{Au$_{2}$(dppe)({\scshape d}-pen)$_{2}$\}$_{2}$] and [Ni\{Rh(apt)$_{3}$\}$_{2}$](NO$_{3}$)$_{2}$ reproduced by the energy position of the satellite structure are shown in Fig.~\ref{Fig.2}(c).
The arctangent-type background estimated from the experimental data originating from the transitions to the continuum states is added to the spectral simulations.
Values of the optimized parameters for the CI cluster-model and effective $d$ electron number ($N_{d}$) and spin states obtained from the simulations are summarized in Table~\ref{Parameter}.
The main and satellite peaks overlap for [Ni$_{2}$\{Au$_{2}$(dppe)({\scshape d}-pen)$_{2}$\}$_{2}$] because of the small ($pd\sigma$). 
The CI cluster-model simulation has reproduced the weak but intrinsic satellite peak position around 861 eV in the photon energy for [Ni\{Rh(apt)$_{3}$\}$_{2}$](NO$_{3}$)$_{2}$.
Note that the slight quantitative inconsistency in the spectral intensity of satellite peak at the $L_{3}$ edge between [Ni\{Rh(apt)$_{3}$\}$_{2}$](NO$_{3}$)$_{2}$ and the [NiS$_{6}$]$^{10-}$ state could be caused by the difference in symmetry.
The spin states of the Ni$^{2+}$ ions obtained from the CI cluster model are qualitatively consistent with those obtained from the magnetic susceptibility for [Ni$_{2}$\{Au$_{2}$(dppe)({\scshape d}-pen)$_{2}$\}$_{2}$] and [Ni\{Rh(apt)$_{3}$\}$_{2}$](NO$_{3}$)$_{2}$~\cite{dppeII,apt}.
We note that $N_{d}$ of [Ni$_{2}$\{Au$_{2}$(dppe)({\scshape d}-pen)$_{2}$\}$_{2}$] is estimated an 8.03 obtained from the fraction of $d^{10-n}$, $d^{11-n}\underbar{L}$, and $d^{12-n}\underbar{L}^{2}$ in Table~\ref{Parameter}, which reflects that the Ni ions are mostly localized.
We have also successfully reproduced the Ni $L_{2,3}$-edge XAS spectra by the HS-Ni$^{2+}$ ionic model for 10$Dq$ = 1.3 eV under the $O_{h}$ symmetry without explicitly taking the hybridizations into account, as shown in Fig.~\ref{Fig.2}(c).
The estimated 10$Dq$, which is {\it renormalized} in the ionic model, can explain the UV and visible absorption spectrum of [Ni$_{2}$\{Au$_{2}$(dppe)({\scshape d}-pen)$_{2}$\}$_{2}$]~\cite{dppeII} combined with the so-called Tanabe-Sugano diagram based on the ionic model~\cite{TSdiagram}, as displayed in Figs.\ref{Fig.2}(d) and (e).
Therefore, the electronic description of [Ni$_{2}$\{Au$_{2}$(dppe)({\scshape d}-pen)$_{2}$\}$_{2}$] is similar to that of Co-based complexes coordinated with {\scshape d}-pen~\cite{Codppe,My paper} for which the explicit hybridization effects are not necessary to describe the $3d$ electronic states. 
On the other hand, ($pd\sigma$) for [Ni\{Rh(apt)$_{3}$\}$_{2}$](NO$_{3}$)$_{2}$ is larger than that for the nickel sulfides and oxides within the cluster models because of the consideration of the final-states effect $\beta$.
In contrast, $\Delta$ is larger than that for nickel sulfides~\cite{NiS2PES}, which may reflect the difference between the molecular, crystal structure and transport properties.

\begin{figure}
\begin{center}
\includegraphics[width=15cm]{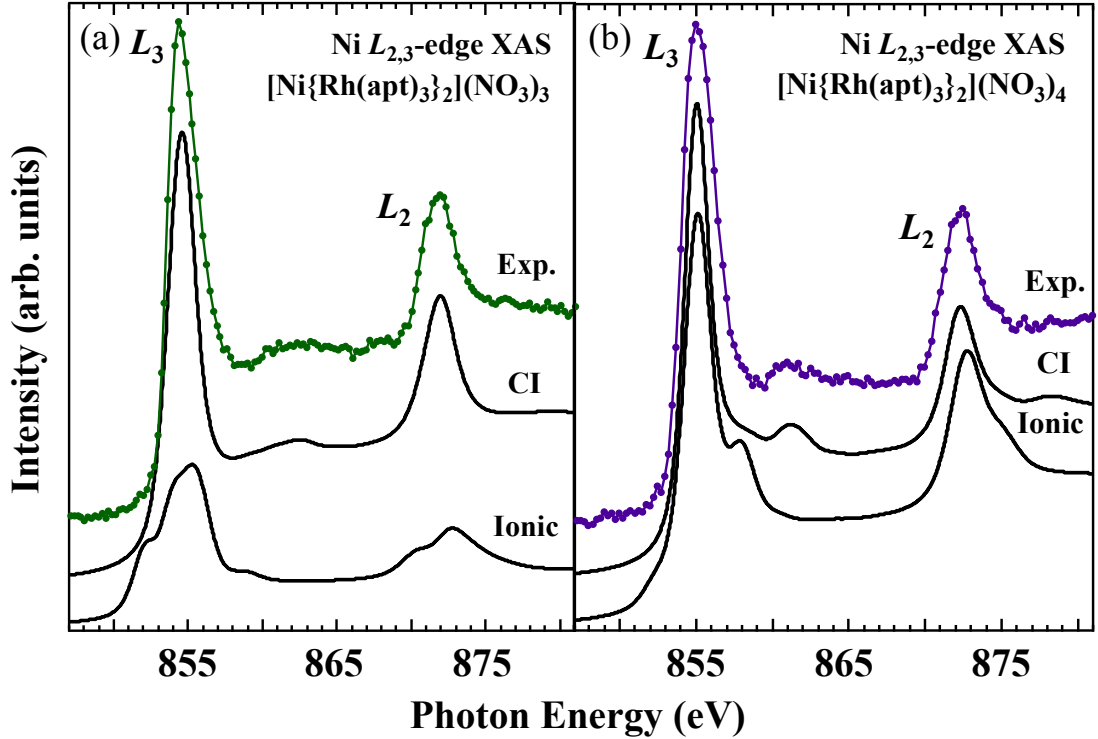}
\end{center}
\vspace{-5mm}
\begin{center}
\caption{(Color online) Comparison between the experimental (color solid lines with circles) and CI cluster-model simulated (black solid lines) Ni $L_{2,3}$-edge XAS spectra of (a) [Ni\{Rh(apt)$_{3}$\}$_{2}$](NO$_{3}$)$_{3}$ and (b) [Ni\{Rh(apt)$_{3}$\}$_{2}$](NO$_{3}$)$_{4}$. All spectra were subtracted by the linear background determined from the raw spectral weight in the region of $h\nu$ = $825-840$ eV. Optimized parameters were summarized in Table~\ref{Parameter}. The ionic-model simulated (black solid lines) Ni $L_{2,3}$-edge XAS spectra with 10$Dq$ = 3.0 eV corresponding to the LS states are also shown.}
\label{Fig.3}
\end{center}
\end{figure}
\subsection{The prominent charge-transfer effects of [Ni\{Rh(apt)$_{3}$\}$_{2}$](NO$_{3}$)$_{n}$}
Hereafter, we focus on the hybridization effects in [Ni\{Rh(apt)$_{3}$\}$_{2}$](NO$_{3}$)$_{n}$.
We show the Ni $L_{2,3}$-edge XAS spectra of [Ni\{Rh(apt)$_{3}$\}$_{2}$](NO$_{3}$)$_{n}$ ($n$ = 3, 4) in Fig.~\ref{Fig.3}.
The $L_{3}$-edge main peak appears to be similar to that of $n$ = 2.
Meanwhile, the satellite peaks away from the main peak at 8.5 eV for $n$ = 3 and 6.0 eV for $n$ = 4 are clearly observed.
The shape of the satellite peak is mutually different, which reflects the difference of the covalency of Ni ions~\cite{apt}.
The simulations of the Ni$^{3+}$ and Ni$^{4+}$ ionic-models with 10$Dq$ = 3.0 eV corresponding to the LS states under the $O_{h}$ symmetry, being consistent with previous calculation results~\cite{LSionicmodel}, are completely contradictory to the experimental XAS results for $n$ = 3, 4 since the simulated $L_{2,3}$-edge main peaks are clearly different and the satellite peaks are not seen in the simulations.
We have confirmed that the multiplet structures are independent on 10$Dq$ within LS states.
Therefore, the hybridization effects between Ni 3$d$ and S 3$p$ orbitals are crucial roles for the electronic states of Ni ions for $n$ = 3, 4.

The simulated Ni $L_{2,3}$-edge XAS spectra of [Ni\{Rh(apt)$_{3}$\}$_{2}$](NO$_{3}$)$_{n}$ ($n$ = 3, 4) by the CI cluster model are also shown in Fig.~\ref{Fig.3}.
The arctangent-type backgrounds estimated from the experimental data are also added to the spectral simulations.
The best-fit parameters $\Delta$ and ($pd\sigma$) to reproduce the energy position of the satellite structure, and obtained $N_{d}$ and spin states for $n$ = 3, 4 are summarized in Table~\ref{Parameter}.
The spin states of the Ni$^{3+}$ and Ni$^{4+}$ ions obtained from the CI cluster model are consistent with those obtained from the magnetic susceptibility for $n$ = 3, 4~\cite{apt}.
The multiplet of $L_{3}$-edge main peak for [Ni\{Rh(apt)$_{3}$\}$_{2}$](NO$_{3}$)$_{3}$ becomes narrower because ($pd\sigma$) is larger value than that for Ni$^{3+}$ oxides $R$NiO$_{3}$ ($R$: 4$f$ rare-earth ion), for which a main peak formed with two distinct components originating from intra-atomic multiplet has been reported, where the typical value of ($pd\sigma$) is 1.5 eV~\cite{RatioIII, RNiO3}.
In addition, the CI cluster-model simulations indicate that the satellite peak of [Ni\{Rh(apt)$_{3}$\}$_{2}$](NO$_{3}$)$_{3}$ has explicitly been developed.
Therefore, the hybridization effects for the Ni$^{3+}$ ions become stronger for sulfur ligands than that for oxygen ligands.
Meanwhile, the simulation for [Ni\{Rh(apt)$_{3}$\}$_{2}$](NO$_{3}$)$_{4}$ can reproduce the experimental spectra with negative $\Delta = -4$ eV.
We have confirmed that the simulated spectra with the positive and small negative $\Delta$ are completely inconsistent with the XAS results because the satellite peak is predicted in the photon energy range of 2$-$3 eV higher than the main peaks by the simulations with such values of $\Delta$.

The negative $\Delta$ has been reported to exist not only in a nominal state of the TM ions such as NaCuO$_{2}$~\cite{NCO1,NCO2} but also the TM ions coordinated with the ligands having a low electronegativity such as Ta$_{2}$NiSe$_{5}$ and NiGa$_{2}$S$_{4}$~\cite{Ta2NiSe5,NiGa2S4}.
The photoemission studies and cluster-model calculations for these inorganic compounds indicate that the ground state is dominantly $d^{m+1}\underbar{L}$ configuration, which induces the charge non-uniformity, the excitonic insulator state~\cite{Ta2NiSe5}, and unusual magnetic interaction~\cite{NiGa2S4}.
Our cluster-model calculation for [Ni\{Rh(apt)$_{3}$\}$_{2}$](NO$_{3}$)$_{4}$ in this study shows that the charge-transferred $d^{7}\underbar{L}$ configuration is more stable than $d^{6}$ configuration with the nominal valency in the ground states as shown in Table~\ref{Parameter}.
We consider that the molecular structure for $n$ = 4 is stable as the [NiS$_{6}$]$^{8-}$ cluster under the octahedral-like symmetry since only less than two electrons are transferred from six sulfur ligands (a total of twelve 3$p$ electrons) to the Ni 3$d$ orbitals.
This tendency of the higher extent of metal-ligand orbital mixing for higher oxidation states of the TM ions have been discussed by the XAS studies for the Ni dithiolene systems~\cite{Ni L-edge, S K-edgeI, S K-edgeII}.

The small and negative $\Delta$ with $(pd\sigma)\gtrsim 1$ eV for $n$ = 3, 4 found in this study also has a key role of physical and chemical properties in the TM complexes.
For instance, quasi one-dimensional Ni$^{3+}$ complex [Ni(chxn)$_{2}$Br]Br$_{2}$ (chxn: 1$R$, 2$R$-cyclohexanediamine) has the gigantic optical nonlinearity~\cite{[Ni(chxn)2Br]Br2} originating from the increase of the transition dipole moment of third-order nonlinear optical response due to the small $\Delta$ (= 1.0 eV) obtained by the angle-resolved photoemission spectroscopy~\cite{[Ni(chxn)2Br]Br2ARPES}.
On the other hand, it has been reported about the strongly mixed configurations of the ligand and Fe 3$d$ states reflecting the delocalized Fe 3$d$ electrons due to the negative $\Delta$ (= $-$1.0 eV) in myoglobin revealed by resonant inelastic soft X-ray scattering~\cite{myoglobin}.
Therefore, we can conclude that the negative $\Delta$ such as [Ni\{Rh(apt)$_{3}$\}$_{2}$](NO$_{3}$)$_{4}$ has a key role of chemical properties such as a redox agent and a catalyst.

\begin{figure}
\begin{center}
\includegraphics[width=12cm]{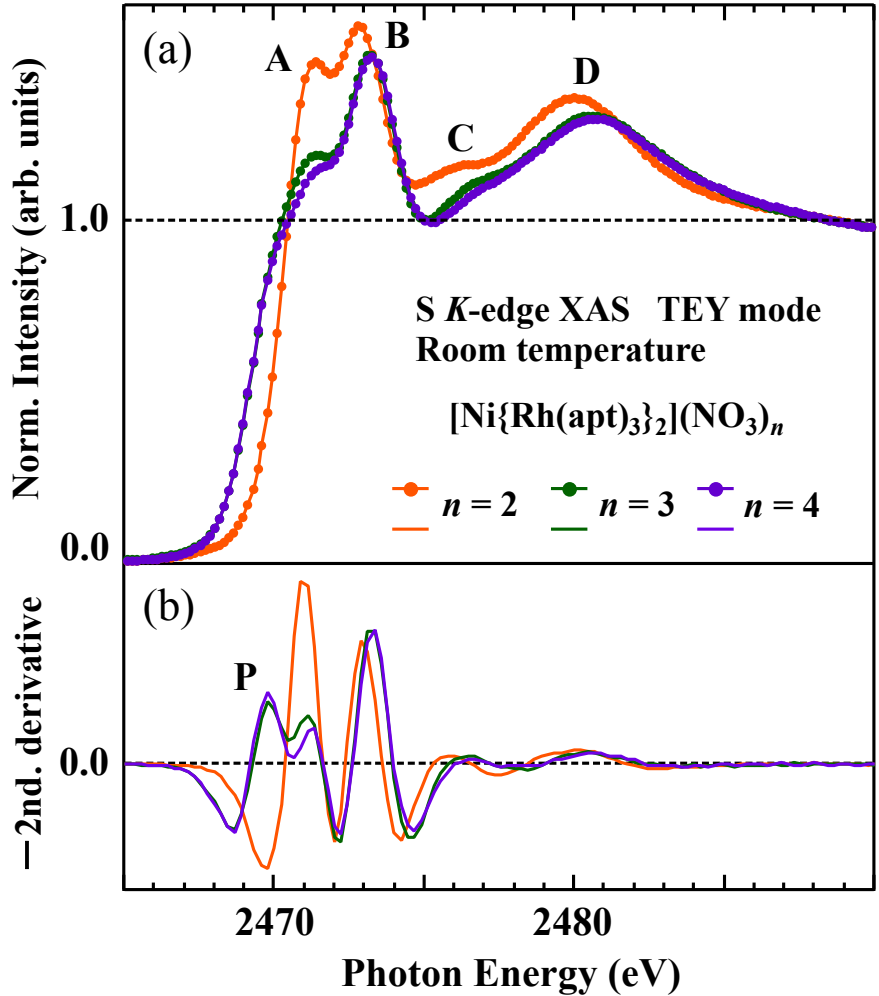}
\end{center}
\vspace{-5mm}
\begin{center}
\caption{(Color online) (a) S $K$-edge XAS spectra of [Ni\{Rh(apt)$_{3}$\}$_{2}$](NO$_{3}$)$_{n}$ at room temperature. All spectra were subtracted by the linear background determined from the raw spectral weight in the region of $h\nu$ = $2450-2460$ eV and normalized by the intensity between 2460 and 2500 eV in the photon energy. The characteristic spectral peaks are labeled as A-D in our experimental spectra. (b) The second derivative of the spectra of [Ni\{Rh(apt)$_{3}$\}$_{2}$](NO$_{3}$)$_{n}$. The peak P corresponds to the singly occupied molecular orbital component as discussed in the text.}
\label{Fig.4}
\end{center}
\end{figure}
To confirm the ligand S 3$p$ electronic states in [Ni\{Rh(apt)$_{3}$\}$_{2}$](NO$_{3}$)$_{n}$, we have also performed the S $K$-edge XAS.
In Fig.~\ref{Fig.4}(a), the characteristic spectral peaks labeled as A-D were observed.
The peaks B ($\sim$2473 eV), C ($\sim$2476 eV), and D ($\sim$2480 eV) shift to higher photon energy with increasing $n$.
The previous S $K$-edge XAS studies for the Ni dithiolene systems, show that the S 1$s$ → C-coordinating S $\pi$ and S 1$s$ → S 4$p$ transitions are located at $\sim$2473.5 eV and higher than 2475.5 eV respectively and depend on the covalency of the Ni ions~\cite{S K-edgeI}.
Therefore, we consider that the peak B is the S 1$s$ → C-coordinating S $\pi$ bonding transition originating from the apt molecules, while the peaks C, and D are the S 1$s$ → S 4$p$ transition.
On the other hand, the peak photon energy of A ($\sim$2471 eV) does not change, in addition, the shoulder structure at $\sim$2470 eV labeled as P has been developed with increasing $n$ revealed by the second derivative for the experimental spectra of [Ni\{Rh(apt)$_{3}$\}$_{2}$](NO$_{3}$)$_{n}$ in Fig.~\ref{Fig.4}(b).
These behaviors have also been reported by the Ni dithiolene systems~\cite{S K-edgeI, S K-edgeII}, where these peaks A and P can be correlated with the Ni 3$d$-S 3$p$ bonding states because of the fact that the electronic states of the Rh ions in [Ni\{Rh(apt)$_{3}$\}$_{2}$](NO$_{3}$)$_{n}$ are independent of $n$, which has been verified by the Rh $L_{3}$-edge XAS spectra~\cite{apt}.

When sulfur atoms have a closed shell structure such as S$^{2-}$ ions, the molecular orbital separates into an occupied and unoccupied orbitals, namely, the highest occupied molecular orbital (HOMO) and lowest unoccupied molecular orbital (LUMO) states can be defined.
In addition, the HOMO would become half-occupied states due to an oxidation or a covalent bonding, so-called the singly occupied molecular orbital (SOMO).
Since the covalent bonding originates from the charge transfer between the TM and ligand orbitals, the SOMO reflects the coexistence of the hybridization effects.
The density functional theory for the Ni dithiolene systems indicates that the photon energy peaks around 2470 and 2471 eV correspond to the SOMO and LUMO states respectively~\cite{S K-edgeII}.
Therefore, we conclude that the peaks A and P correspond to the LUMO and SOMO for [Ni\{Rh(apt)$_{3}$\}$_{2}$](NO$_{3}$)$_{n}$.

It is well known that the decrease of the interatomic distance leads to the increase of ($pd\sigma$) due to the greater overlap of the 3$d$ and neighboring $p$ states.
Indeed, the average bonding length between Ni and S ions becomes small from 2.418 \AA\ for $n$ = 2 to 2.343 \AA\ for $n$ = 3, which has been obtained by X-ray diffraction~\cite{apt}, being consistent with the increasing value of ($pd\sigma$) obtained by the CI cluster model with increasing $n$, as shown in Table~\ref{Parameter}.
Meanwhile, ($pd\sigma$) of $n$ = 4 is smaller than that of $n$ = 2, 3 due to the different nominal ionic configuration $d^{m}$.
On the other hand, the effective transfer integral $\sqrt{\alpha}T_{\sigma}$ = 1.91 eV between $d^{7}\underbar{L}$ and $d^{8}\underbar{L}^{2}$ states for $n$ = 4 is comparable to that between $d^{7}$ and $d^{8}\underbar{L}$ states for $n$ = 3.
Such a comparison is meaningful since the $d^{7}\underbar{L}$ configuration is dominant for $n$ = 4 due to the negative $\Delta$.

On the other hand, $\Delta$ decreases with increasing metal ion valence and decreasing the electronegativity of ligands for inorganic compounds~\cite{NiS2PES,Nidihalides}.
In the case of [Ni\{Rh(apt)$_{3}$\}$_{2}$](NO$_{3}$)$_{n}$, $\Delta$ also decreases with increasing the Ni ion covalency since the 3$d$ level are lowered as the $n$ increases.
As shown in Table~\ref{Parameter}, the fraction of $d^{10-n}$, $d^{11-n}\underbar{L}$, and $d^{12-n}\underbar{L}^{2}$ obtained by the CI cluster-model simulations for Ni $L_{2,3}$-edge XAS spectra shows that the $d^{11-n}\underbar{L}$ configuration becomes more stable than the $d^{10-n}$ configuration at the boundary between $n$ = 2, 3, leading the large $N_{d}$.
The increasing stability of the $d^{11-n}\underbar{L}$ configuration is attributed to the decreasing $\Delta$ rather than the increasing ($pd\sigma$).
The intensity ratio of the SOMO to the LUMO obtained from the S $K$-edge XAS spectra is 0.30 for $n$ = 2 and 0.70 for $n$ = 3 respectively, corresponding to the increasing fraction of the $d^{11-n}\underbar{L}$ configuration qualitatively.
On the other hand, the $d^{12-n}\underbar{L}^{2}$ configuration is more stabilized from the $d^{11-n}\underbar{L}$ configuration for $n$ = 4, on going to the negative $\Delta$.
$N_{d}$ for $n$ = 4 indicates that the electrons that transfer between the Ni and S ions are more than one, being different from those of $n$ = 2, 3.
Although the intensity ratio of the SOMO to the LUMO of $n$ = 4 is similar 0.72 to that of $n$ = 3, these ratios can be correlated with the total fraction of the $d^{11-n}\underbar{L}$ and $d^{12-n}\underbar{L}^{2}$ configurations, which qualitatively corresponds to the fact that the absorbance of the ligand to metal charge transfer components increases with increasing $n$ by the visible ultraviolet absorption spectra~\cite{apt}.
Therefore, we conclude that the strength of the charge-transfer effects is dominantly due to $\Delta$, not ($pd\sigma$) within [Ni\{Rh(apt)$_{3}$\}$_{2}$](NO$_{3}$)$_{n}$.

As demonstrated above, it has been found that the prominent charge-transfer effects highly deviated from the nominal valences of the Ni ions are seen in [Ni\{Rh(apt)$_3$\}$_2$](NO$_3$)$_n$ ($n = 3,4$) due to the Ni $3d$-S $3p$ hybridizations. 
In other words, the electrons with the highest energy seem to be delocalized in the vicinity of the Ni ions.  
On the other hand, the change of the {\it valence} of the sulfur ions is few, at most 0.27 even for $n=4$, 
since the coordination number is six. 
Considering the facts that the sulfur ions are bound to the nearest carbon sites with single bond and that the Rh electronic states change hardly with $n$~\cite{apt}, we can judge that a possible further charge transfer from the sulfur ions is not likely within the apt molecules. 
Therefore, the degree of the charge delocalization centered at the Ni ions is at most up to the nearest sulfur sites 
whereas the gross electronic structure of the apt molecules is rather independent of $n$. 
These may lead to the interconvertible nature of [Ni\{Rh(apt)$_3$\}$_2$](NO$_3$)$_n$ through the redox reactions with keeping the coordination geometry as well as the stability of the high oxidation state of Ni$^{4+}$.  
Our findings imply the potentials of the apt complexes coordinated with the sulfur ions as new functional materials.

\section{Conclusion}

In conclusion, we have studied the local 3$d$ electronic states of the sulfur coordinating Ni based complexes [Ni$\{$Rh(apt)$_{3}\}_{2}$](NO$_{3}$)$_{n}$ and [Ni$_{2}$\{Au$_{2}$(dppe)({\scshape d}-pen)$_{2}$\}$_{2}$] by means of X-ray absorption spectroscopy and CI cluster-model simulations.
The hybridization effects between the Ni 3$d$ and S 3$p$ orbitals depend on the sulfur coordination number and $n$.
The Ni $L_{2,3}$-edge XAS spectral simulations by the cluster model under the $O_{h}$ symmetry provides that the hybridization effects are stronger for [Ni$\{$Rh(apt)$_{3}\}_{2}$](NO$_{3}$)$_{n}$ than for inorganic compounds.
The observation of the singly occupied molecular orbitals from the S $K$-edge XAS spectra supports the experimental and simulation results for the Ni $L_{2,3}$-edge XAS. 
These lead to the prominent charge-transfer effects on the Ni sites for $n=3,4$, which may have a key role for the interconvertible nature as well as the stability of the high oxidization state of the Ni ions in [Ni$\{$Rh(apt)$_{3}\}_{2}$](NO$_{3}$)$_{n}$. 

\section*{Acknowledgments}

We thank Y. Kanai, T. Kita, M. Murata, M. Yamada, and T. Ohta for supporting the experiments and H. Fujiwara for flutiful discussions.
The XAS study was supported by Project for Creation of Research Platforms and Sharing of Advanced Research Infrastructure, Japan (No. R1545, S16006).
K. Yamagami and M. Kouno were supported by the Program for Leading Graduate Schools ``Interactive Materials Science Cadet Program''.

\end{document}